\documentclass[journal,subeqn,caption]{IEEEtran}
\usepackage[singlelinecheck=false,labelsep=period,font=footnotesize]{caption}
\pdfoutput=1
\usepackage{mathtools}
\usepackage{bbold}
\usepackage{tabularx}
\usepackage{cite}
\usepackage{float}
\usepackage{multicol}
\usepackage{multirow}
\usepackage{hyperref}
\usepackage{graphicx}
\usepackage{amsfonts}
\usepackage{amsmath}
\usepackage{color}
\usepackage{epstopdf}
\usepackage{caption}
\usepackage{subcaption}
\usepackage{mathtools, nccmath}
\usepackage{multirow}
\usepackage{bbm}
\usepackage{marginnote}
\usepackage{tikz}
\usepackage{enumitem}
\usepackage{dblfloatfix}
\usepackage{comment}
\usepackage{eurosym}
\usepackage{makecell}
\usepackage[switch]{lineno}
\usetikzlibrary{arrows,decorations.pathmorphing,backgrounds,fit,positioning,shapes.symbols,chains}
\usepackage{colortbl}

 \ifCLASSINFOpdf
\else
\fi
\newenvironment{ldescription}[1]
  {\begin{list}{}%
   {\renewcommand\makelabel[1]{##1\hfill}%
   \settowidth\labelwidth{\makelabel{#1}}%
   \setlength\leftmargin{\labelwidth}
   \addtolength\leftmargin{\labelsep}}}
  {\end{list}}

\begin{document}
\title{EV-based Smart E-mobility System -- Part II: Formulation and Case Study}
\author{I. Pavi\'c, \textit{Student Member}, \textit{IEEE}, 
\thanks{Ivan Pavi\'c, Hrvoje Pand\v{z}i\'c and Tomislav Capuder are with the University of Zagreb Faculty of Electrical Engineering and Computing, Zagreb HR-10000, Croatia (e-mails: \href{mailto:ivan.pavic@fer.hr}{ivan.pavic@fer.hr},  \href{mailto:hrvoje.pandzic@fer.hr}{hrvoje.pandzic@fer.hr}, \href{mailto:tomislav.capuder@fer.hr}{tomislav.capuder@fer.hr}).}
%
H. Pand\v{z}i\'c, \textit{Senior Member}, \textit{IEEE},
T. Capuder, \textit{Member}, \textit{IEEE}
}
%
%
\maketitle

\begin{abstract}
This is the second paper of a two-paper series on smart e-mobility. The concept defined in the first paper is further elaborated and mathematically formulated in here. A case study is conducted to evaluate the differences between the proposed electric-vehicle-based and the existing charging-station-based e-mobility systems. Each of the four issues identified in the first paper are individually examined and omission of corresponding constraints are analyzed and quantified.
\end{abstract}


\section{Models}

To demonstrate the points from the former companion paper, models of both the EV-based (EVBA) and the CS-based aggregator (EVCA) are formulated in the following subsections and evaluated in the case study presented in Section \ref{sec:res}.

\subsection{Nomenclature} \label{sec:nomen}
\subsubsection{Abbreviations}
\begin{ldescription}{$xxxxxx$}
\item [BMS] Battery management system.
\item [CC-CV] Constant-current-constant-voltage.
\item [CP] Charging point.
\item [CS] Charging station.
\item [DOD] Depth-of-discharge.
\item [EV] Electric vehicle.
\item [EVBA] Electric vehicle battery aggregator.
\item [EVCA] Electric vehicle charge aggregator.
\item [LIB] Lithium-ion battery.
\item [OBC] On-board charger.
\item [OF] Objective function.
\item [SOE] State-of-energy.
\item [V2G] Vehicle-to-grid.
\end{ldescription}
\subsubsection{Sets and Indices}
\begin{ldescription}{$xxxxxx$}
\item [$\mathcal{CP}$] Set of charging points, indexed by $cp$.
\item [$\mathcal{T}$] Set of time steps, indexed by $t$.
\item [$\mathcal{V}$] Set of vehicles, indexed by $v$.
\end{ldescription}
\subsubsection{Input parameters}
\begin{ldescription}{$xxxxxxxx$}
\item [$C^{\text{BAT}}_{v}$] Capital battery cost of vehicle $v$ (\EUR).
\item [$C^{\text{CP\_FCH}}_{v,t,cp}$] Charging point fee for fast chargers at charging point $cp$ (\EUR/kWh).
\item [$C^{\text{CP\_SCH}}_{v,t,cp}$] Charging point fee for slow chargers at charging point $cp$ (\EUR/kWh).
\item [$C^{\text{EP}}_{t}$] Electricity price during period $t$ (\EUR/kWh).
\item [$C^{\text{G\_FCH}}_{v,t,cp}$] Grid tariff for fast chargers at charging point $cp$ (\EUR/kWh).
\item [$C^{\text{G\_SCH}}_{v,t,cp}$] Grid tariff for slow chargers at charging point $cp$ (\EUR/kWh).
\item [$CAP^{\text{BAT}}_{v}$] Battery capacity of vehicle $v$ (kWh).
\item [$D^{\text{BAT}}_{1}$] Fixed battery degradation coefficient for higher values of depth-of-discharge.
\item [$D^{\text{BAT}}_{2}$] Variable battery degradation coefficient (based on discharged energy) for higher values of depth-of-discharge.
\item [$D^{\text{BAT}}_{3}$] Variable battery degradation coefficient (based on depth-of-discharge) for higher values of depth-of-discharge.
\item [$D^{\text{BAT}}_{4}$] Variable battery degradation coefficient (based on discharged energy) for lower values of depth-of-discharge.
\item [$E^{\text{CP\_MAX}}_{cp}$] Maximum energy limit of charging point $cp$ during one time step (kWh).
\item [$E^{\text{FCH\_MAX}}$] Maximum energy limit of fast charging point during one time step (kWh).
\item [$E^{\text{OBC\_MAX}}_{v}$] Maximum energy limit of OBC of vehicle $v$ during one time step (kWh).
\item [$E^{\text{RUN}}_{v,t}$] Energy consumed for mobility purposes of vehicle $v$ during time step $t$.
\item [$SOE^{\text{ARR}}_{v,cp}$] Anticipated SOE at time of arrival at $cp$ of vehicle $v$ in a CS-based system.
\item [$SOE^{\text{CV}}_{v}$] SOE curve breaking point between CC and CV charging phases of vehicle $v$ (\%).
\item [$SOE^{\text{DEP}}_{v,cp}$] Anticipated SOE at time of departure from $cp$ of vehicle $v$ in a CS-based system.
\item [$SOE^{\text{MIN}}_{v}$] Minimum allowed SOE of vehicle $v$ (\%).
\item [$SOE^{\text{MAX}}_{v}$] Maximum allowed SOE of vehicle $v$ (\%).
\item [$SOE^{\text{0}}_{v}$] Initial SOE of vehicle $v$ (\%).
\item [$T^\mathrm{ARR}_{v,cp}$] Time step when vehicle $v$ arrives at charging point $cp$ in a CS-based system.
\item [$T^\mathrm{DEP}_{v,cp}$] Time step when vehicle $v$ departs from charging point $cp$ in a CS-based system.
\item [$T^\mathrm{OFF}_{v,cp}$] Set of time steps when vehicle $v$ when vehicle $v$ is disconnected from charging point $cp$ in a CS-based system.
\item [$T^\mathrm{ON}_{v,cp}$]
Set of time steps when vehicle $v$ is connected to charging point $cp$ in a CS-based system.
\item [$\eta^{\text{DCH}}$] EV V2G discharging efficiency.
\item [$\eta^{\text{FCH}}$] EV fast charging efficiency.
\item [$\eta^{\text{RUN}}$] EV mobility discharging efficiency.
\item [$\eta^{\text{SCH}}$] EV slow charging efficiency.
\item [$\mathbb{1}_{v,t,cp}$] Matrix indicating whether vehicle $v$ is connected to charging point $cp$ at time step $t$.
\end{ldescription}
\subsubsection{Variables}
\begin{ldescription}{$xxxxxxxx$}
\item [$c^{\text{DEG}}_{v,t}$] Degradation cost of vehicle $v$ at time $t$ (\EUR).
\item [$c^{\text{EV}}$] Overall cost of charging all EVs (\EUR).
\item [$e^{\text{DCH}}_{v,t}$] Energy discharged from vehicle $v$ at time $t$ (kWh).
\item [$e^{\text{FCH}}_{v,t}$] Energy fast charged to vehicle $v$ at time $t$ (kWh).
\item [$e^{\text{SCH}}_{v,t}$] Energy slow charged to vehicle $v$ at time $t$ (kWh).
\item [$soe^{\text{EV}}_{v,t}$] State-of-energy of vehicle $v$ at time $t$ (kWh).
\end{ldescription}
\subsection{Mathematical Formulation of an EV-based Aggregator} \label{sec:EVmod}

Objective function minimizes the total EV charging costs:

\begin{gather} 
\min_{\Xi^{\text{OF}}} c^{\text{EV}} =\sum_{v\in\mathcal{V}} \Big[ \sum_{t\in\mathcal{T}} \Big( e^{\text{SCH}}_{v,t} \cdot \left( C^{\text{EP}}_{t} + C^{\text{G\_SCH}}_{v,t,cp} + C^{\text{CP\_SCH}}_{v,t,cp} \right) 
\nonumber \\
- e^{\text{DCH}}_{v,t} \cdot C^{\text{EP}}_{t}  + c^{\text{DEG}}_{v,t} 
\nonumber \\
+ 
e^{\text{FCH}}_{v,t} \cdot \left( C^{\text{EP}}_{t} + C^{\text{G\_FCH}} + C^{\text{CP\_FCH}} \right)  \Big) \Big]. \label{OF}
\end{gather}

The first row in Eq.~\eqref{OF} corresponds to payments due to EV charging at slow chargers, where $e^{\text{SCH}}_{v,t}$ is charged energy, $C^{\text{EP}}_{t}$ is energy price, $C^{\text{G\_SCH}}_{v,t,cp}$ is the grid fee for slow chargers\footnote{Slow chargers refer to AC chargers, i.e the ones that require OBC to convert alternating to direct current.} and $C^{\text{CP\_SCH}}_{v,t,cp}$ is the CS fee. The second row represents EV discharging income and cost of degradation, where $e^{\text{DCH}}_{v,t}$ is the amount of discharged energy, $C^{\text{EP}}_{t}$ is V2G revenue and $c^{\text{DEG}}_{v,t}$ battery degradation cost. The third row captures payments due to EV charging at fast chargers\footnote{Fast chargers refer to DC chargers, i.e the ones that convert alternating to direct current and circumvent the OBC. Therefore, the OBC capacity is not relevant when using fast chargers.}, where $e^{\text{FCH}}_{v,t}$ is the amount of charged energy, $C^{\text{G\_FCH}}$ is the grid fee for fast chargers, and $C^{\text{CP\_FCH}}$ is the fast CS fee. EV slow charger charging fees depend on the type of charger, e.g. this fee is zero for home chargers. On the other hand, EV fast charging is modeled using only one fast charging type and cost. 

Charging/discharging energy constraints are:
\begin{gather}
e^{\text{SCH}}_{v,t}, e^{\text{DCH}}_{v,t}, e^{\text{FCH}}_{v,t} \geq 0 \quad \forall v\in \mathcal{V}, t\in \mathcal{T}; \label{e_SDF}
\\
e^{\text{SCH}}_{v,t} \leq \sum_{cp\in\mathcal{CP}}\mathbb{1}_{v,t,cp}\cdot E^{\text{CP\_MAX}}_{cp} \quad \forall v\in \mathcal{V}, t\in \mathcal{T}; \label{e_SCH_CP}
\\
e^{\text{DCH}}_{v,t} \leq \sum_{cp\in\mathcal{CP}}\mathbb{1}_{v,t,cp}\cdot E^{\text{CP\_MAX}}_{cp} \quad \forall v\in \mathcal{V}, t\in \mathcal{T}; \label{e_DCH_CP}
\\
e^{\text{SCH}}_{v,t} \leq E^{\text{OBC\_MAX}}_{v} \quad \forall v\in \mathcal{V}, t\in \mathcal{T}; \label{e_SCH_OBC}
\\
e^{\text{DCH}}_{v,t} \leq E^{\text{OBC\_MAX}}_{v} \quad \forall v\in \mathcal{V}, t\in \mathcal{T}; \label{e_DCH_OBC}
\\
e^{\text{SCH}}_{v,t} \leq E^{\text{OBC\_MAX}}_{v} \cdot \frac{1-soe^{\text{EV}}_{v,t}}{1-SOE^{\text{CV}}_{v} \cdot CAP^{\text{BAT}}_{v}} 
\nonumber \\
\forall  v\in \mathcal{V}, t\in \mathcal{T}; \label{e_SCH_CV}
\\
e^{\text{FCH}}_{v,t} \leq \sum_{cp\in\mathcal{CP}}\mathbb{1}_{v,t,cp}\cdot E^{\text{FCH\_MAX}} \quad v\in \mathcal{V}, t\in \mathcal{T}. \label{e_FCH_CP}
\end{gather}

Constraint~\eqref{e_SDF} imposes nonnegativity on all energy variables. Constraints~\eqref{e_SCH_CP} and~\eqref{e_DCH_CP} limit the energy charged/discharged at slow CSs based on the mapping parameter $\mathbb{1}_{v,t,cp}$ that determines which EV is connected to which CP at each time step. As the EVs move between different CPs, maximum charging power depends on index $cp$. OBC limits on EV slow charging and discharging are imposed by constraints~\eqref{e_SCH_OBC} and~\eqref{e_DCH_OBC}, respectively. The OBC power capacity $E^{\text{OBC\_MAX}}_{v}$ depends only on the EV type. Constraint~\eqref{e_SCH_CV} additionally constrains the OBC charging power at high state-of-energy (SOE) due to inherent nature of the li-ion battery (LIB) charging process consisting of the constant-current (CC) and the constant-voltage (CV) part. Parameter $SOE^{\text{CV}}$ is empirically obtained and indicates SOE value (in percentage) at which the constant voltage phase starts. More information on this formulation can be found in \cite{Vagropoulos2015} and \cite{Pandzic2018}. Finally, the fast charging power limit $E^{\text{FCH\_MAX}}$ is imposed by constraint~\eqref{e_FCH_CP}.

LIB degradation is calculated as follows:
\begin{gather}
c^{\text{DEG}}_{v,t} \geq C^{\text{BAT}}_{v} \cdot (D^{\text{BAT}}_{1} + D^{\text{BAT}}_{2} \cdot \frac {e^{\text{DCH}}_{v,t}} {CAP^{\text{BAT}}_{v}} \cdot 100 
\nonumber \\
+ D^{\text{BAT}}_{3} \cdot \frac{1-soe^{\text{EV}}_{v,t}} {CAP^{\text{BAT}}_{v}} \cdot 100 ) \quad \forall v \in \mathcal{V},t\in \mathcal{T}; \label{DEG_1}\\
c^{\text{DEG}}_{v,t} \geq C^{\text{BAT}}_{v} \cdot (D^{\text{BAT}}_{4} \cdot \frac {e^{\text{DCH}}_{v,t}} {CAP^{\text{BAT}}_{v}} \cdot 100) 
\nonumber \\
\quad \forall v \in \mathcal{V},t \in \mathcal{T}. \label{DEG_2}
\end{gather}

LIB degradation depends on four main variables: charging/discharging current, voltage, temperature and cell balance. In most LIB applications the last two variables are kept at optimal operating point by a dedicated battery management system (BMS) and they can be left out of the degradation model. During slow AC charging the currents are rather low (up to 0.2C\footnote{C-rate is the ratio of the charging (or discharging) power and battery energy capacity.}) and their impact on degradation is negligible. Thus, the only variable that must be taken into account is voltage, which is closely related to SOE, thus constraints~\eqref{soe_MIN} and~\eqref{soe_MAX} keep the voltage within the allowed range. In order to consider degradation, a penalization cost is introduced as in \cite{Ortega-Vazquez2014}, but in a linearized form in order to avoid binary variables  \cite{RecaldeMelo2018}. Geometric surface of the linearized degradation cost is modeled by constraint~\eqref{DEG_1}, which includes two variables: discharged energy and depth-of-discharge (DOD = 1 -- SOE). Constraint~\eqref{DEG_2} is an additional geometric surface binding at higher values of SOE when surface from eq.~\eqref{DEG_1} goes to zero or becomes negative. Constraint~\eqref{DEG_2} depends only on discharged energy. Parameters $D_{1-4}$ are obtained using the best-fit option applied to LIB degradation data (life-cycle loss vs. DOD) from \cite{Ecker2014}.

Energy balance constraints are:
\begin{gather}
soe^{\text{EV}}_{v,t} = soe^{\text{EV}}_{v,t-1} + e^{\text{SCH}}_{v,t} \cdot \eta^{\text{SCH}} 
- e^{\text{DCH}}_{v,t} / \eta^{\text{DCH}} 
\nonumber\\
- E^{\text{RUN}}_{v,t} / \eta^{\text{RUN}} 
+ e^{\text{FCH}}_{v,t} \cdot \eta^{\text{FCH}}
 \quad \forall v \in \mathcal{V},t\in \mathcal{T};
 \label{SOE_c}\\
soe^{\text{EV}}_{v,t} \geq SOE^{\text{MIN}}_{v} \cdot CAP^{\text{BAT}}_{v} \quad \forall v \in \mathcal{V},t\in \mathcal{T}; \label{soe_MIN}\\
soe^{\text{EV}}_{v,t} \leq SOE^{\text{MAX}}_{v} \cdot CAP^{\text{BAT}}_{v} \quad \forall v \in \mathcal{V},t\in \mathcal{T}; \label{soe_MAX}\\
soe^{\text{EV}}_{v,t} \geq SOE^{\text{0}}_{v} \cdot CAP^{\text{BAT}}_{v} \quad \forall v \in \mathcal{V},t=24; \label{soe_24}
\end{gather}

Eq.~\eqref{SOE_c} is the main energy balance equation calculated for each vehicle $v$ and time step $t$. Energy accumulated during the current time step must be equal to the energy accumulated in the previous time step plus the energy withdrawn from the grid via slow or fast charging points and minus the energy discharged for motion or back into the grid. In the first time step the term $soe^{\text{EV}}_{v,t-1}$ is substituted with $SOE^{\text{0}}_v$, which corresponds to energy stored in vehicle $v$ before the first time step. Constraints~\eqref{soe_MIN} and~\eqref{soe_MAX} limit the battery capacity of each EV, while constraint~\eqref{soe_24} sets the minimum SOE in the last time step.

\subsection{Mathematical Formulation of a CS-based Aggregator} \label{sec:CSmod}

Mathematical model of the CS-based aggregator is:

\begin{gather}
\text{min (1)}\nonumber\\
\text{subject to}\qquad\qquad\qquad\qquad\qquad\qquad\qquad\qquad\qquad\qquad\nonumber\\
(2)-(10), (12)-(14)\nonumber\\
soe^{\text{EV}}_{v,t} = soe^{\text{EV}}_{v,t-1} + e^{\text{SCH}}_{v,t} \cdot \eta^{\text{SCH}} 
- e^{\text{DCH}}_{v,t} / \eta^{\text{DCH}} 
\nonumber\\
+ e^{\text{FCH}}_{v,t} \cdot \eta^{\text{FCH}}
 \quad \forall v \in \mathcal{V},t\in T^\text{ON}_{v,cp}; \label{e_SOE4V}
\\
soe^{\text{EV}}_{v,t} = SOE^\text{ARR}_{v,cp},\quad\forall v\in \mathcal{V},t=T^\text{ARR}_{v,cp},cp\in \mathcal{CP}; \label{SOE_ARR}
\\
soe^{\text{EV}}_{v,t} \geq  SOE^\text{DEP}_{v,cp}, \quad \forall v\in \mathcal{V},t=T^\text{DEP}_{v,cp}, cp\in \mathcal{CP}.
\label{SOE_DEP}
\end{gather}

It contains all constraints as the EV-based aggregator model except for~\eqref{SOE_c}, which is replaced with constraints \eqref{e_SOE4V}--\eqref{SOE_DEP}. Energy balance constraint~\eqref{e_SOE4V} does not include energy discharge for driving as it only tracks the EVs when they are connected to a CP. Hence the time domain in eq.~\eqref{e_SOE4V} is $T^\text{ON}_{v,cp}$. Eqs.~\eqref{SOE_ARR} and~~\eqref{SOE_DEP} are used to set the anticipated SOE at arrival and required SOE at departure from each CP.

\section{Results and Discussion} \label{sec:res}
\emph{Issues 1 \& 2} (insufficient information on EV behavior at other CSs and inability to transfer flexibility between CSs) are observed together as they both depend on the EVs' daily SOE curve. \emph{Issues 3 \& 4} (insufficient power constraints and incomplete costs) are addressed individually and only for the EVBA case, as their repercussions are the same for both models. 
In this test case we use the illustrative example with three EVs and three CSs from the Part I paper. 

\subsection{Input Data}\label{sec:input}

\begin{figure}[t]
    \centering
    \includegraphics[height=\linewidth/2, width=\linewidth]{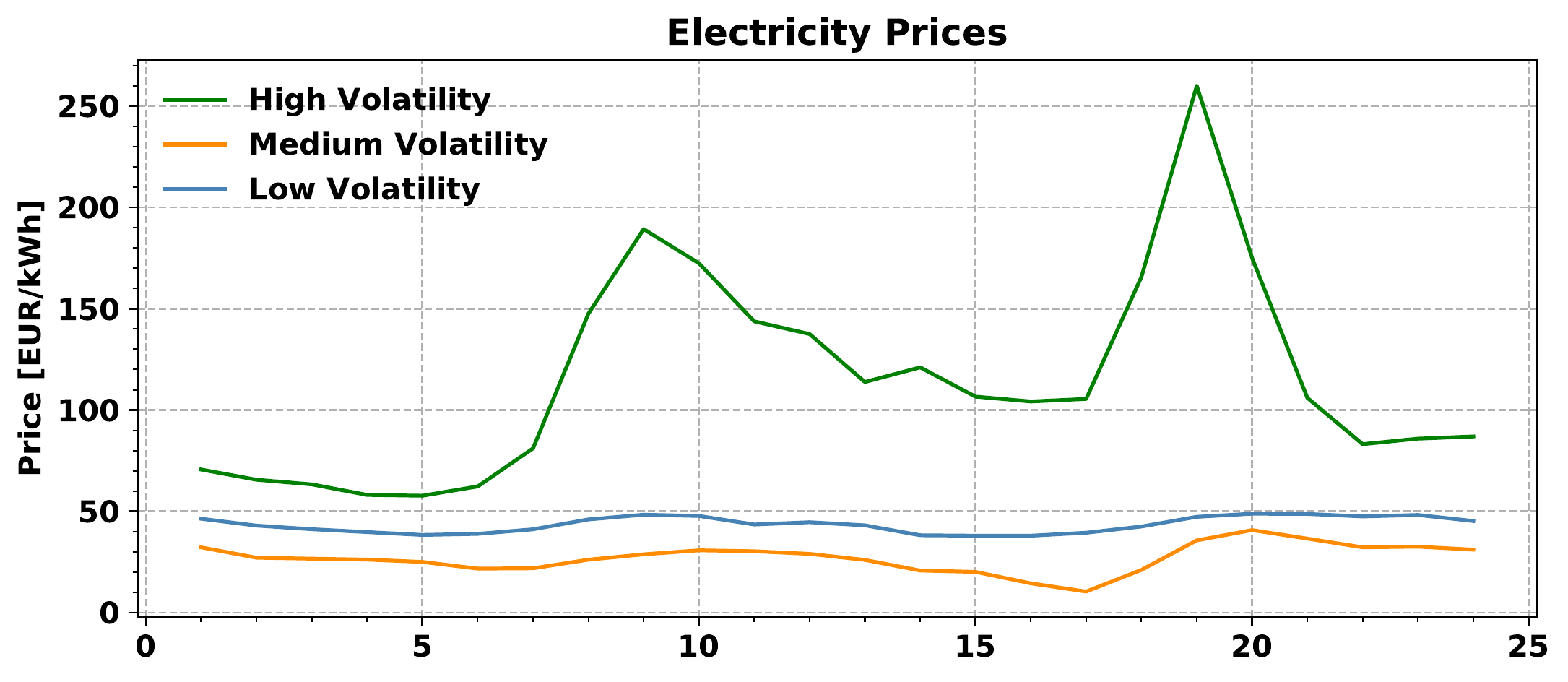}
    \caption{Three scenarios of electricity prices from EPEX}
    \label{fig:prices}
\end{figure}

The proposed model resembles a price taker scheme where an aggregator forecasts prices in order to efficiently submit its energy bids in the market. Although both the prices, driving activity and times of arrival and departure from CPs are stochastic parameters, we consider all parameters deterministic for better demonstration of optimality of both formulations, as well as quantification of the resulting schedules. 

We use historic energy prices data for year 2018 from EPEX power exchange in France. Three sets are used resembling high, medium and low volatility of electricity prices, as shown in Figure \ref{fig:prices}. The high-volatility prices date from Nov. 21, medium from March 11, and low from June 30. Each charger type has different grid and charger tariff fee, as listed in Table \ref{tab:cpdata}. All grid fees are modeled using a two-tariff system: night and day, and the fees are aligned with the ones in \cite{Grave2016}. Grid fees represent both transmission and distribution fees, while charger fees are used to retrieve investment and cover for operation and maintenance costs of the charging infrastructure. Generally, higher charger power results in lower grid fees, but higher charger fees. Charger fees are obtained from real fast charging fees in \cite{Zap} and \cite{Jungle} reduced by average energy price and grid tariff fees and scaled based on investment cost to match the corresponding charger type. The investment costs of chargers are from \cite{Invest}.

\begin{table}[b]
 \centering
 \captionsetup{justification=centering}
\caption{Charger point (CP) data (kW and \EUR{/kW})}\label{tab:cpdata}
\setlength{\tabcolsep}{3pt} 
 \begin{tabular}{c c c c c c} 
 \hline
 \multirow{2}{*}{CP Type} & \multirow{2}{*}{Description} & Power  & Grid Low  & Grid High & CP Tariff \\
  &  & (kW)  & (\EUR{/kW})  & (\EUR{/kW}) & (\EUR{/kW}) \\ 
 \hline\hline
 1 & Home &  4 & 0.02284 & 0.047040 & 0.004\\ 
 \hline
 2 & Work &  8 & 0.016120 & 0.033600 & 0.0183  \\ 
 \hline
 3 & Leisure &  12 & 0.016120 & 0.033600 & 0.03 \\ 
 \hline
 4 & DC Fast &  100 & 0.010750 & 0.022840 & 0.2 \\ 
 \hline
\end{tabular}
\end{table}

\begin{figure*}[!b]
    \centering
    \includegraphics[width=\linewidth]{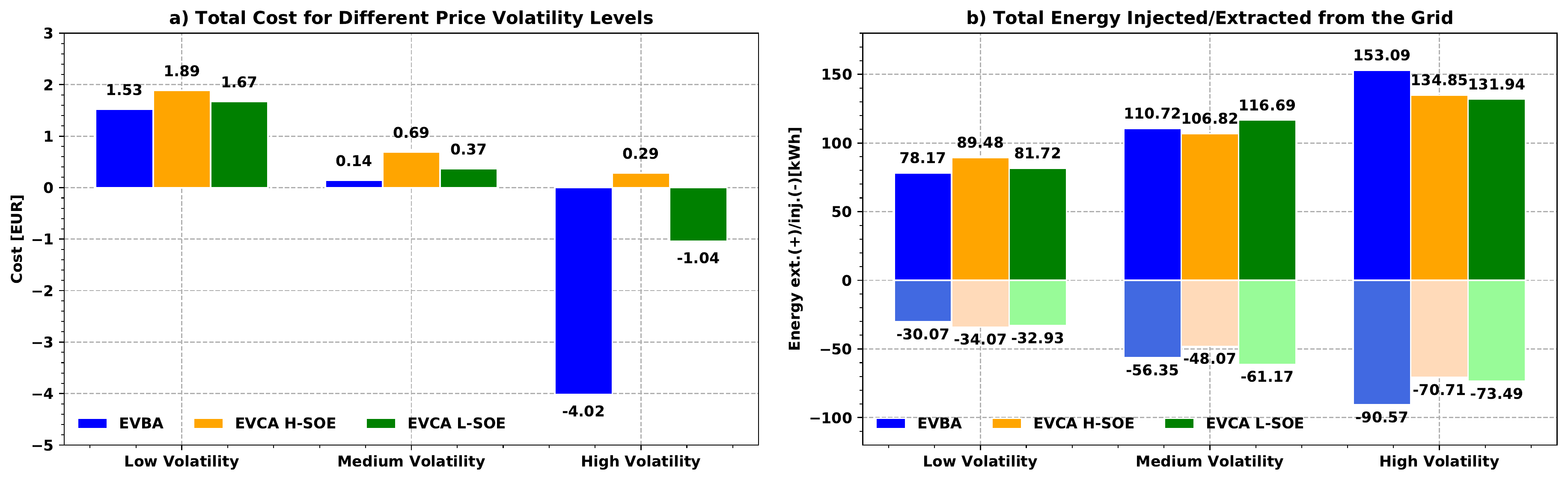}
    \caption{Results related to \emph{Issues 1 \& 2}, showing total charging costs and energy injection/extraction for all three EVs}
    \label{fig:Issue12}
\end{figure*}

\begin{figure*}[!b]
    \centering
    \includegraphics[width=\linewidth]{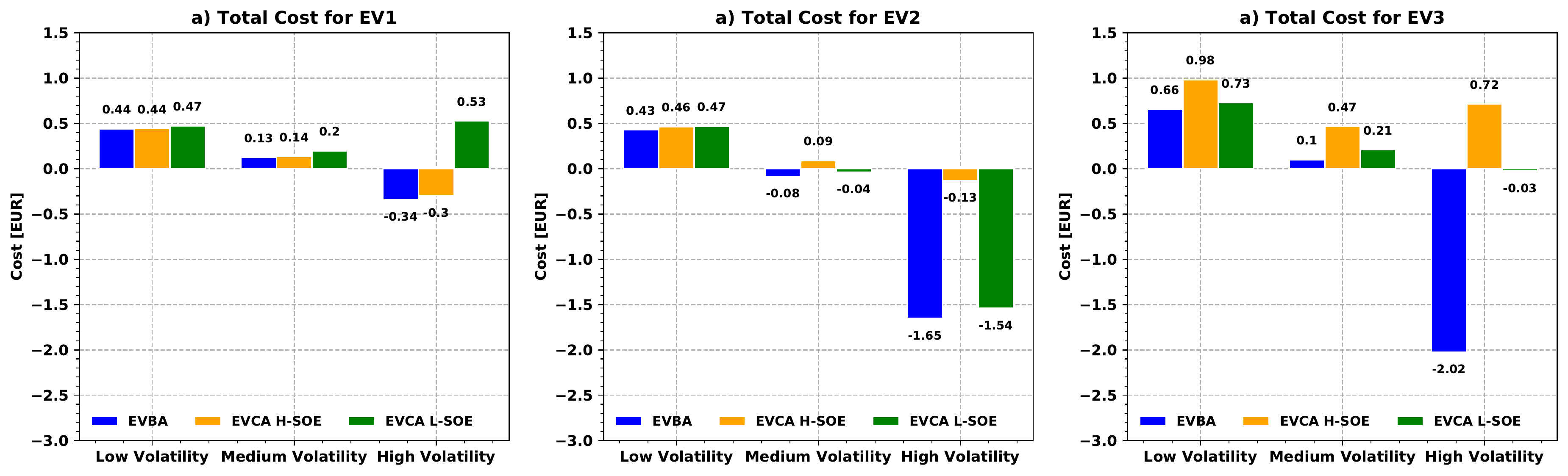}
    \caption{Results related to \emph{Issues 1 \& 2}, showing total charging costs for each EV individually}
    \label{fig:CostEV123}
\end{figure*}

\begin{figure*}[!b]
    \centering
    \includegraphics[width=\linewidth]{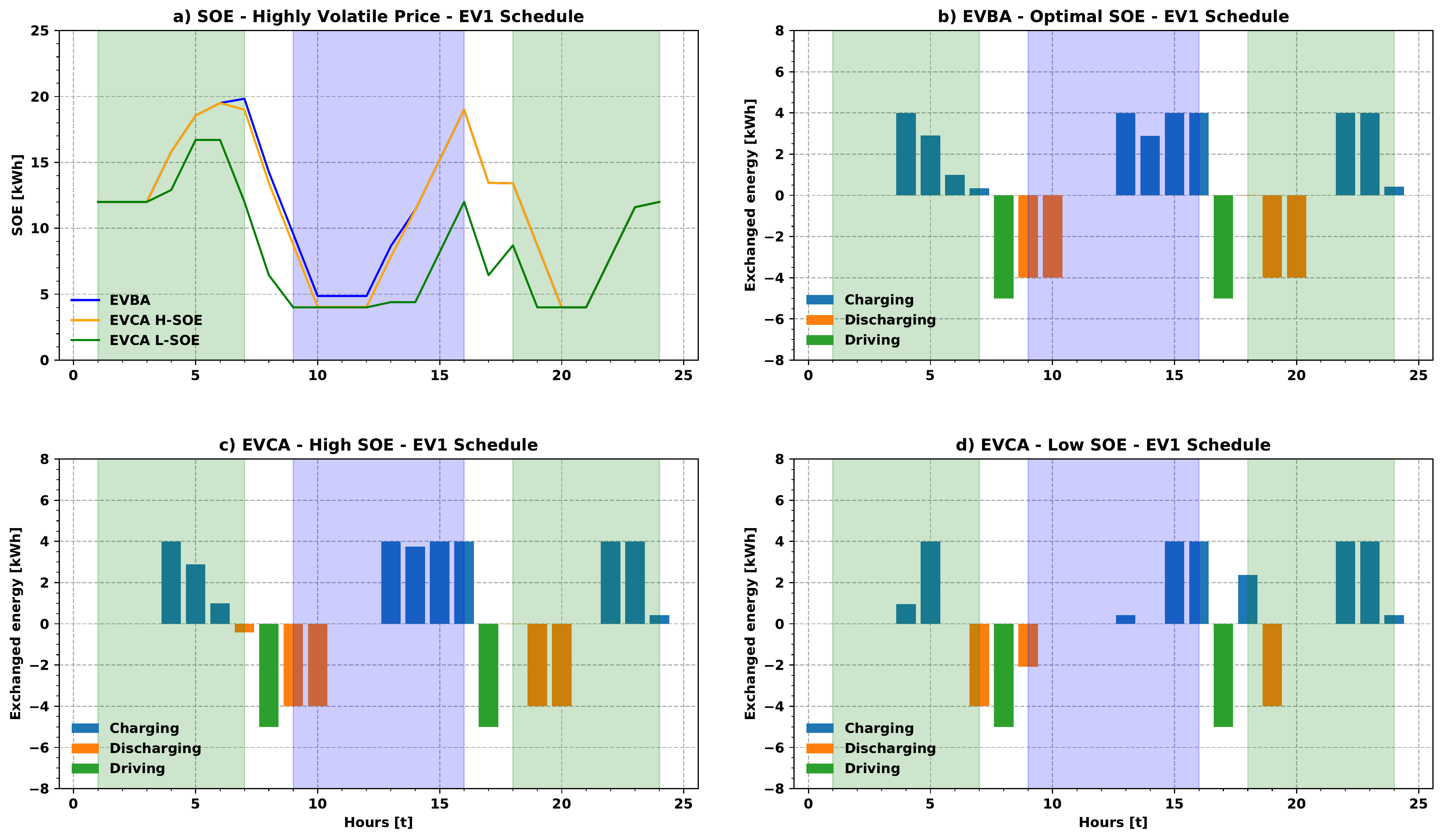}
    \caption{Results related to \emph{Issues 1 \& 2}, EV1 schedules for the highly volatile price scenario}
    \label{fig:Issue12EV1}
\end{figure*}

\begin{figure*}[!b]
    \centering
    \includegraphics[width=\linewidth]{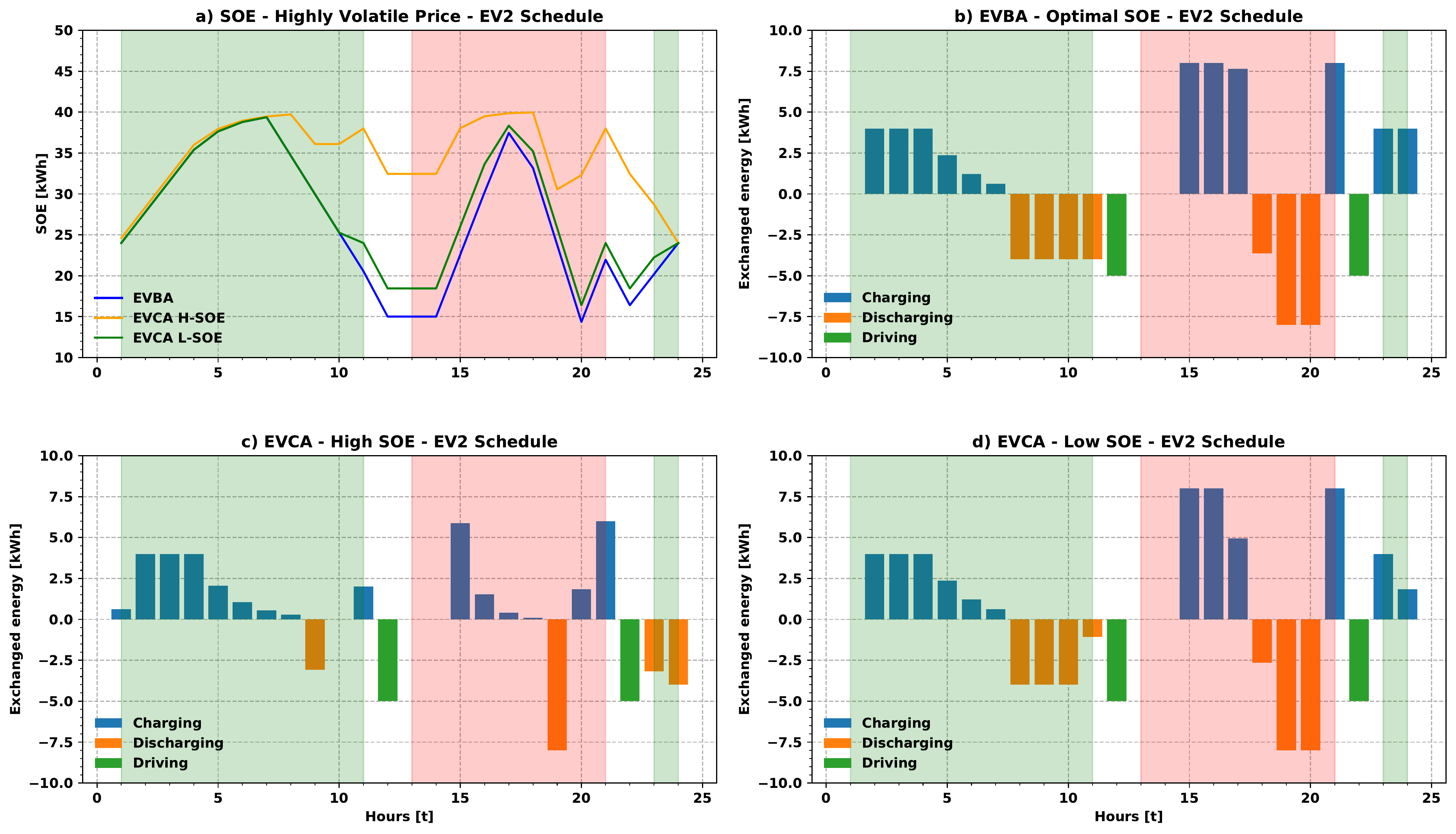}
    \caption{Results related to \emph{Issues 1 \& 2}, EV2 schedules for the highly volatile price scenario}
    \label{fig:Issue12EV2}
\end{figure*}

\begin{figure*}[!b]
    \centering
    \includegraphics[width=\linewidth]{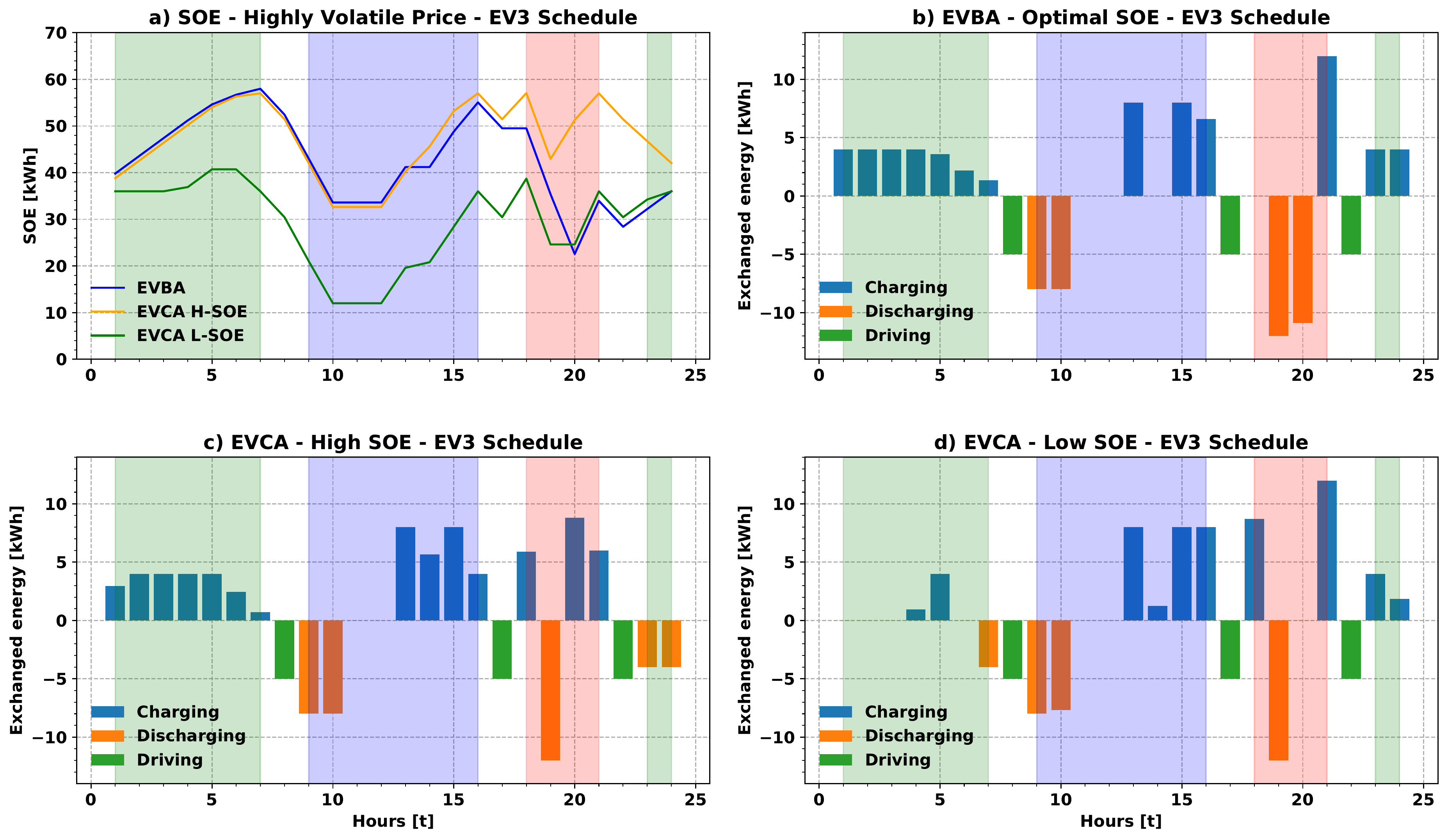}
    \caption{Results related to \emph{Issues 1 \& 2}, EV3 schedules for the highly volatile price scenario}
    \label{fig:Issue12EV3}
\end{figure*}

Efficiencies used in this paper are as follows: slow charging $\eta^{\text{SCH}}=0.95$, discharging for driving $\eta^{\text{RUN}}=0.90$, discharging to drive $\eta^{\text{DCH}}=0.85$, and fast charging $\eta^{\text{FCH}}=0.80$. SOE parameters used for all EVs are following: $SOE^{\text{MAX}}=1$, $SOE^{\text{MIN}}=0.2$, $SOE^{\text{CV}}=0.8$, and $SOE^{\text{0}}=0.6$. Battery capacities are 20 kWh for EV1, 40 kWh for EV2 and 60 kWh for EV3. Battery degradation parameters are: $D^{\text{BAT}}_{1}=-0.3429$, $D^{\text{BAT}}_{2}=0.03403$, $D^{\text{BAT}}_{3}=0.004287$, and $D^{\text{BAT}}_{4}=0.008317$. 

To highlight \emph{Issues 1 \& 2} in the EVCA model, two different values of $SOE_{v,cp}^\mathrm{DEP}$ are used. The first one corresponds to a conservative driver who sets the SOE before every trip to at least 95\% (nearly full), and we name this model \textit{high-SOE}. The second one corresponds to a risk-prone driver willing to earn more for providing flexibility at an expanse of its EV range. This person sets the SOE before every trip to at least 60\%. We name this model \textit{low-SOE}. Note that most models in literature assume a conservative driver who always desires (nearly) full battery at departure. 

\begin{figure*}[!b]
    \centering
    \includegraphics[width=\linewidth]{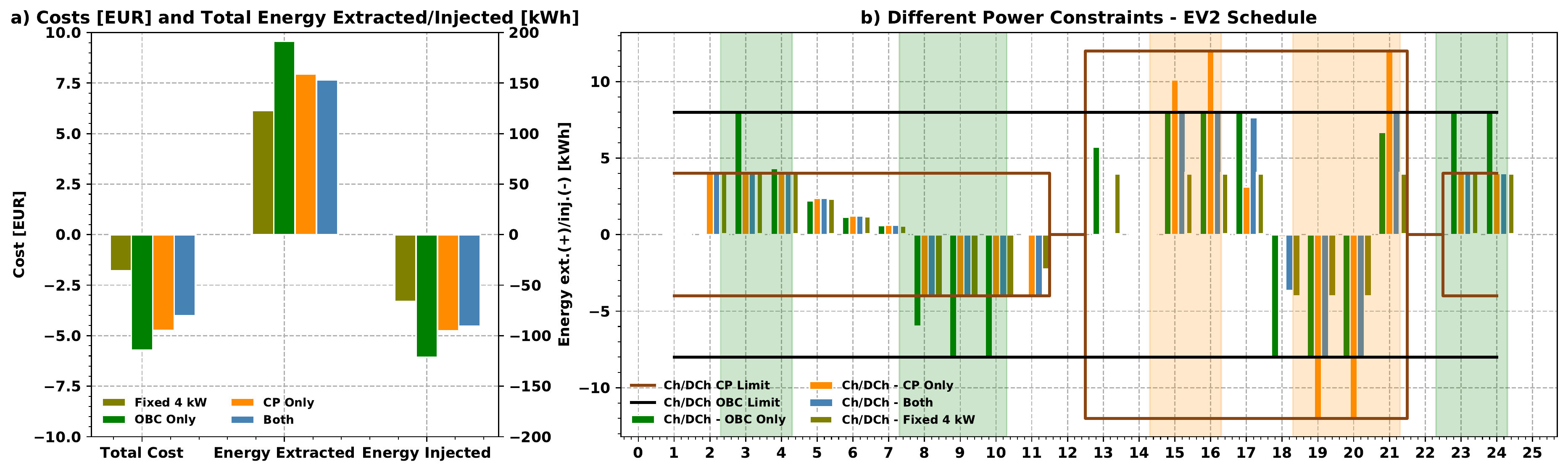}
    \caption{Results related to \emph{Issue 3}}
    \label{fig:Issue3}
\end{figure*}

\begin{figure*}[!b]
    \centering
    \includegraphics[width=\linewidth]{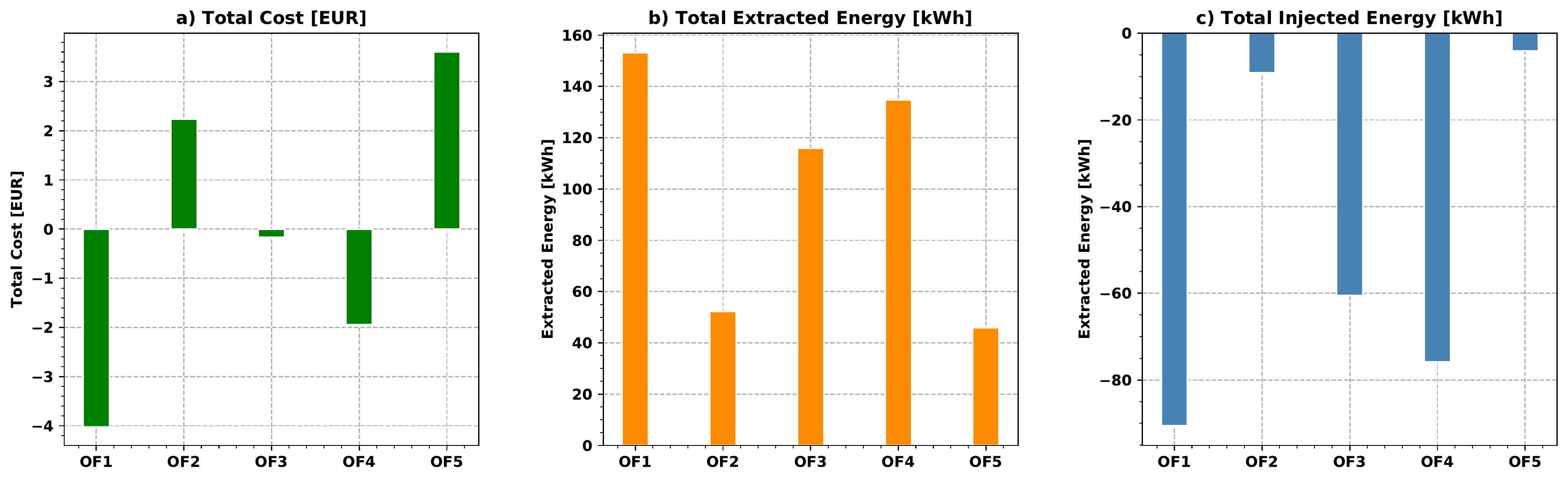}
    \caption{Results related to \emph{Issue 4} (OF1 -- only cost of electricity; OF2: cost of electricity + degradation cost; OF3: cost of electricity + grid tariff; OF4: cost of electricity + CS tariff, OF5: electricity cost + degradation cost + grid tariff + CS tariff)}
    \label{fig:Issue4}
\end{figure*}

\subsection{Issues 1 \& 2}

The results related to \emph{Issues 1 \& 2} are displayed in Figures \ref{fig:Issue12}--\ref{fig:Issue12EV3}. 
Results in Figure \ref{fig:Issue12}a indicate that in total, i.e. combined for all three EVs, the EVBA model results in the lowest charging costs for all price volatility scenarios, followed by the EVCA low-SOE, while the worst results are achieved for the EVCA high-SOE model. Detailed individual EV costs are shown in Figure \ref{fig:CostEV123}, where the EVBA model provides the cheapest solution for all three EVs over all price volatility scenarios, while the two EVCA cases alternate in terms of the quality of the solution. For EV1, the high-SOE case is always a better option, while for EV2 the low-SOE case is a better option for all price scenarios. For EV3 however, in low-volatility price scenario the high-SOE case yields better results, while in medium- and high-volatility scenarios the low-SOE case performs better. The reason for EVBA superiority over the EVCA models are twofold: i) in the EVCA models the EVs are often charged at high prices and ii) their energy arbitrage opportunities are reduced due to strict SOE requirements. Generally, all three models discharge most energy in the high-volatility price scenario as such scenario favors arbitrage, as can be seen in Figure \ref{fig:Issue12}b. In the low-volatility scenario the EVBA model is the least aggressive in V2G mode, but in the high-volatility scenario it discharges the most energy. In all price-volatility scenarios the EVBA model observes price differences during the whole day and adjusts its discharging schedule accordingly. On the other hand, in EVCA models the CSs are blind to prices outside of the timeframe when an EV is connected to them and they need to adjust their discharging quantities to keep the departing SOE at the minimum allowed level. This happens even if this discharge incurs higher recharging costs at subsequent CSs. 

In general, higher price volatility yields lower costs in all three cases. However, the EVBA model is able to better monetize flexibility over the day and the charging costs reduce drastically as the price volatility increases (EV2 generates profit already in medium-volatility price scenario). This is highly related to \emph{Issue 2} (transfer of flexibility between CSs). Since the EVBA model observes EVs throughout the day, it can schedule optimal amount of discharging when prices are high allowing the EVs to drive to another CSs with sufficient SOE. 

\emph{Issue 1} (problems with SOE prediction at EV arrival) are analyzed in details in Figures \ref{fig:Issue12EV1}--\ref{fig:Issue12EV3} for the highly volatile price scenario. In all three figures the periods when EVs are parked at CSs, are shaded in the respective CS color. 
In case of EV1 and highly volatile prices, the first driving period precedes the periods of high prices. In the EVBA model, EV1 charges before the first trip and discharges after, as shown in Figure \ref{fig:Issue12EV1}b. It recharges before the second trip (during the low-price hours 13-16) and again discharges at the next CS. It charges for the last time at the end of the day at low prices. A similar schedule is obtained with the EVCA high-SOE model. However, CS1 is oblivious to the low prices in the afternoon and slightly discharges EV1 in hour 7, as opposed to the EVBA model that charges EV1 in hour 7 (compare Figures \ref{fig:Issue12EV1}b and \ref{fig:Issue12EV1}c). To make up for this lack of energy, the high-SOE EVCA model needs to charge more energy in hour 14 than the EVBA model. This is suboptimal since the price in hour 14 is much higher than in hour 7. The charging quantities in all the other hours are the same. Graph in Figure \ref{fig:Issue12EV1}d indicates that the EVCA low-SOE model behaves quite differently than the other two. Since the CS before the first trip only satisfies the EV's desired SOE of 60\% at the departure and at the same time minimizes costs of EV charging only at this CS, it significantly discharges EV1 before the first trip. When prices are highest, after the first trip, EV1 discharges much less energy than in the other two cases due to lower SOE after the trip. Before the second trip, EV1 is again charged only to satisfy the desired SOE at the next departure time, and therefore has less energy to be discharged after the second trip (compare hours 19 and 20). Considering the SOE graphs and  charging schedules from Figure \ref{fig:Issue12EV1}, the conclusion is that the EVCA high-SOE model performs much closer to the optimal EVBA model than the ECVA-low model. 

In the case of EV2 and highly volatile prices (Figure \ref{fig:Issue12EV2}) the first driving period takes place after the periods of high prices. In the EVBA model, whose charging schedule is shown in Figure \ref{fig:Issue12EV2}b, EV2 charges early in the morning and discharges before the first trip taking advantage of peaking prices in hours 8-11. It fully recharges after the first trip (hours 15-17) to be able to fully discharge during hours 18-20. Energy for the second trip is charged just before the trip, in hour 21, at very low cost. The required SOE is achieved by charging EV2 after the final trip at low cost (hours 23 and 24). Comparison of the EVBA charging schedule and the low-SOE EVCA schedule in Figure \ref{fig:Issue12EV2}d, as well as the corresponding daily SOE curves in Figure \ref{fig:Issue12EV2}a, indicates that the low-SOE EVCA model behaves quite similar to the optimal EVBA model. The only differences are as follows:
\begin{itemize}
    \item The EVCA low-SOE model discharges less energy in hour 11 as it requires at least 60\% of SOE at departure. 
    \item Due to higher SOE, the EVCA low-SOE model requires less charging in hour 17. Since the electricity price in hour 11 is much higher than in hour 17, this model overlooked an arbitrage opportunity between those hours.
    \item Again, due to 60\% required SOE, the EVCA low-SOE model discharges less energy in hour 18.
    \item Due to higher SOE, the EVCA low-SOE model requires less charging in hour 24. Again, it did not exercise arbitrage between hours 18 and 24 due to a required SOE level at departure.
\end{itemize}

The results of the high-SOE EVCA are shown in Figure \ref{fig:Issue12EV2}c. This model does not take advantage of discharging at higher prices due to a more constrained SOE requirement at departure and thus results in much worse solution than the EVCA high-SOE model. For instance, instead of discharging in hours 8-11 as the EVBA and EVCA low-SOE models, the EVCA high-SOE model is, due to the departing SOE restriction, only able to perform partial discharge in hour 9. This repeats again in the evening hours when the EVCA high-SOE model is only able to perform discharge in hour 19, instead of hours 18-20. As a consequence, the EVCA high-SOE model is left with a lot of energy stored in the late evening hours. This energy is partially discharged in the last two hours of the day, but at a relatively low profit.

The EV3 case for the highly volatile prices is shown in Figure \ref{fig:Issue12EV3}. In the EVBA model (Figure \ref{fig:Issue12EV3}b), EV3 charges before the first trip and discharges after it to take advantage of peak price hours 9 and 10. It recharges before the second trip to be able to discharge again after the trip, thus performing arbitrage. It again recharges before and after the third trip to meet the required SOE at the end of the day. Graphs in Figure \ref{fig:Issue12EV3}a indicate that optimal EVBA case is similar to the high-SOE EVCA case during the morning and the daytime, but during the evening it resembles the low-SOE case. The morning charging period at CS1 (green area) ends at hour 7, when the high-SOE EVCA model charges EV3 to 95\%, as required by this model. This is quite similar to the optimal EVBA model, which charges EV3 only to a slightly higher SOE. At CS2 (blue area), the high-SOE model charges the EV again to 95\%, while the EVBA model charges it slightly below that value. The major difference occurs in the evening hours at CS3 (red area), where the high-SOE EVCA model again charges EV3 to 95\% of its SOE, while the EVBA model charges it to only 33 kWh in hour 21. This demonstrates the negative effect of constraint on the departure SOE in the high-SOE EVCA model. EV3 is thus required to charge instead of discharge at very high prices. Consequently, after the final trip it has more energy then required by the end-of-day SOE constraint and CS1 (green area) discharges it, but at a low gain, in hours 23 and 24.

The EVCA low-SOE case schedules EV3 quite differently before the first and second trips. It does not charge as much energy since the required SOE before the trips is only 60\%. This enables it to perform arbitrage at CS1 and discharge a part of the energy in hour 7 just before the trip (Figure \ref{fig:Issue12EV3}c). Since hours 9 and 10 are peak-price hours, it discharges more energy and charges again in hours 13-16 at lower prices. It again performs arbitrage in hours 19 and 21, but with much lower energy volume than the EVBA model.  

Based on the conducted analysis of the EV behavior, we derive the following conclusions:
\begin{enumerate}
\item for EV1, the high-SOE EVCA model is close to the optimal EVBA model;
\item for EV2, the low-SOE EVCA model is close to the optimal EVBA model,
\item in the case of EV3, the high-SOE EVCA model is close to the optimal EVBA solution until evening, but during the evening and night the low-SOE EVCA case becomes more similar to the optimal EVBA solution.
\end{enumerate}

Therefore, without the EVBA optimization model there is no way to decide what is the best required SOE at the time of departure to maximally transfer flexibility and utilize daily energy arbitrage.  

\subsection{Issue 3}

To analyze \emph{Issue 3} (insufficient power constraints), we examine the results of the EVBA model with highly volatile prices using four different sets of power constraints. First, we use fixed power constraint of 4 kW throughout the day. Second and third sets of constraints use only OBC and CP power constraints, respectively. The fourth set of constraints uses both the OBC and CP power constraints. 

As shown in Figure \ref{fig:Issue3}a, the minimum expected costs are obtained when using only OBC power constraint, followed by the CP-only power constraint, then both power constraints, while the highest cost is obtained for a fixed 4 kW power constraint. This is a direct result of energy arbitrage volumes shown in the same chart. In order to verify feasibility of the obtained charging schedules, Figure \ref{fig:Issue3}b shows the exceeded OBC and CP limits. The green shaded areas indicate that the injected/extracted power exceeds the CP limit, while the orange shaded areas indicate the surpassed CP limit. The CP power limit is exceeded in hours 3, 8-10, 23 and 24 by the OBC-only case as the OBC rated power is higher than the CS1 rated power. On the other hand, the OBC power capacity is exceeded in hours 15, 16, 19-21 by the CP-only case as the OBC capacity is lower than the CP capacity during those hours. Cases with fixed 4 kW power constraint and inclusion of both the OBC and CP power constraints never exceed the power limits. Therefore, the cases with only OBC and only CP power constraints provide higher revenues only at first sight. However, their real-time operation cannot be physically carried out and they would suffer from additional balancing costs not included in Figure \ref{fig:Issue3}a. On the contrary, if EVs are too constrained, as in the case with fixed 4 kW power limit, the EV charging schedule is overconstrained, which diminishes the arbitrage opportunities. This brings us to conclusion that considering both the OBC and CP power constraints results in optimal solution.

\subsection{Issue 4}

From mathematical perspective, \emph{Issue 4} (incomplete costs) deals with different terms in the objective function. Figure \ref{fig:Issue4} shows that adding the cost terms usually omitted in the existing literature significantly reduces the attractiveness of energy arbitrage. Five objective functions (OF) with different elements are observed: 

\begin{enumerate}
\item OF1: base case with only the cost of electricity,
\item OF2: cost of electricity and battery degradation costs,
\item OF3: cost of electricity and grid tariff, 
\item OF4: cost of electricity and CS tariff,
\item OF5: all the costs, including cost of electricity, battery degradation costs, grid tariff and CS tariff. 
\end{enumerate}

The graph in Figure \ref{fig:Issue4}a shows that the total cost rises from -4 \EUR in the electricity-only case to 3.6 \EUR in the case with all relevant costs included, which makes a huge difference in the EV charging economics. The main factor are degradation costs (OF2 value is 2.2 \EUR), while the lowest impact has the CS tariff (OF4 value is -1.9 \EUR). 

The overall costs are in direct relation with the volume of arbitrage as the spread in the price between the purchased and is the sold electricity needs to cover for additional costs of battery degradation and tariffs. Therefore, OF5 results in the least charged energy, followed by OF2, as shown in Figure \ref{fig:Issue4}b. With respect to this, total discharged energy reduced from 90,57 kWh in the OF1 case to a mere 4,07 kWh in the all-costs case, as shown in Figure \ref{fig:Issue4}c.  

\section{Conclusion} \label{sec:concl}
The paper demonstrated important findings in the field of smart e-mobility outlined in the Part I paper. The commonly observed e-mobility system where CSs takes the leading role in electricity markets yields sub-optimal results for the EV owners. The proposed e-mobility system where EVs take the leading role in electricity markets proved to be much more economically attractive for EV owners. This is especially the case when volatility of electricity prices is high. In such case the EV-based model results in 3.87 times lower overall costs for the three observed EVs than the CS-based models. Opposed to the EV-based model, the analyzed CS-based models cannot accurately anticipate the optimal arriving and departing SOE and cannot exchange flexibility among CSs. 

Also, the paper showed that insufficiently modeled constraints and costs can steer the scheduling results in a wrong direction leading to infeasible charging/discharging bids and higher actual operating costs. Analysis of accurate power constraints points out the value of higher installed power capacities both for OBC and external CS equipment. 

Further research will focus on uncertainty in EV-based models and participation of an EVBA in ancillary services markets.

\bibliography{sample} 

\begin{thebibliography}{1}
\providecommand{\url}[1]{#1}
\csname url@samestyle\endcsname
\providecommand{\newblock}{\relax}
\providecommand{\bibinfo}[2]{#2}
\providecommand{\BIBentrySTDinterwordspacing}{\spaceskip=0pt\relax}
\providecommand{\BIBentryALTinterwordstretchfactor}{4}
\providecommand{\BIBentryALTinterwordspacing}{\spaceskip=\fontdimen2\font plus
\BIBentryALTinterwordstretchfactor\fontdimen3\font minus
  \fontdimen4\font\relax}
\providecommand{\BIBforeignlanguage}[2]{{%
\expandafter\ifx\csname l@#1\endcsname\relax
\typeout{** WARNING: IEEEtran.bst: No hyphenation pattern has been}%
\typeout{** loaded for the language `#1'. Using the pattern for}%
\typeout{** the default language instead.}%
\else
\language=\csname l@#1\endcsname
\fi
#2}}
\providecommand{\BIBdecl}{\relax}
\BIBdecl

\bibitem{Vagropoulos2015}
S.~I. {Vagropoulos}, D.~K. {Kyriazidis}, and A.~G. {Bakirtzis}, ``Real-time
  charging management framework for electric vehicle aggregators in a market
  environment,'' \emph{IEEE Transactions on Smart Grid}, vol.~7, no.~2, pp.
  948--957, March 2016.

\bibitem{Pandzic2018}
H.~{Pand\v{z}i\'{c}} and V.~{Bobanac}, ``An accurate charging model of battery
  energy storage,'' \emph{IEEE Transactions on Power Systems}, vol.~34, no.~2,
  pp. 1416--1426, March 2019.

\bibitem{Ortega-Vazquez2014}
M.~A. Ortega-Vazquez, ``{Optimal scheduling of electric vehicle charging and
  vehicle-to-grid services at household level including battery degradation and
  price uncertainty},'' \emph{IET Generation, Transmission {\&} Distribution},
  vol.~8, no.~6, pp. 1007--1016, June 2014.

\bibitem{RecaldeMelo2018}
D.~F. {Recalde Melo}, A.~Trippe, H.~B. Gooi, and T.~Massier, ``{Robust Electric
  Vehicle Aggregation for Ancillary Service Provision Considering Battery
  Aging},'' \emph{IEEE Transactions on Smart Grid}, vol.~9, no.~3, pp.
  1728--1738, May 2018.

\bibitem{Ecker2014}
M.~Ecker, N.~Nieto, S.~K{\"{a}}bitz, J.~Schmalstieg, H.~Blanke, A.~Warnecke,
  and D.~U. Sauer, ``{Calendar and cycle life study of Li(NiMnCo)O2-based 18650
  lithium-ion batteries},'' \emph{Journal of Power Sources}, vol. 248, pp.
  839--851, Feb. 2014.

\bibitem{Grave2016}
K.~Grave, B.~Breitschopf, J.~Ordonez, J.~Wachsmuth, S.~Boeve, M.~Smith,
  T.~Schubert, N.~Friedrichsen, A.~Herbst, K.~Eckartz, M.~Pudlik, M.~Bons,
  M.~Ragwitz, and J.~Schleich, ``{PRICES AND COSTS OF EU ENERGY Final
  Report},'' European Commission, Tech. Rep., 2016.

\bibitem{Zap}
\BIBentryALTinterwordspacing
``{EV rapid charge cost comparison - Zap-Map}.'' [Online]. Available:
  \url{https://www.zap-map.com/ev-rapid-charge-cost-comparison/}
\BIBentrySTDinterwordspacing

\bibitem{Jungle}
\BIBentryALTinterwordspacing
``{The jungle of charge tariffs in the Netherlands - Amsterdam University of
  Applied Sciences}.'' [Online]. Available:
  \url{http://www.idolaad.com/shared-content/blog/rick-wolbertus/2016/charge-tarrifs.html}
\BIBentrySTDinterwordspacing

\bibitem{Invest}
\BIBentryALTinterwordspacing
``{RMI: What's the true cost of EV charging stations? | GreenBiz}.'' [Online].
  Available:
  \url{https://www.greenbiz.com/blog/2014/05/07/rmi-whats-true-cost-ev-charging-stations}
\BIBentrySTDinterwordspacing

\end{thebibliography}

\end{document}